\documentclass[twocolumn,showpacs, pre]{revtex4}

\usepackage{amsmath}
\usepackage{amssymb} 
\usepackage{amsfonts}
\usepackage{graphicx}
\usepackage[hyperindex,breaklinks]{hyperref}
\usepackage{theorem}
                 \def\d{\delta}     
            \def\h{\eta}     
        \def\l{\lambda}                  
\def\n{\nu}           \def\p{\pi}           
\def\s{\sigma}        \def\t{\tau}                
          \def\ps{\psi}              
                    
        \def\L{\Lambda}       
                           
\def\th{\vartheta}            
                  \def\o{\omega}

\def\bfe{{\bf e}} 
\def\pp{{\bf p}}\def\xx{{\bf x}}
 
\def\yy{{\bf y}}\def\kk{{\bf k}}

\def\hw{{\widehat w}}

\def\hp{{\widehat \ps}}

\def\be{\begin{equation}}    \def\ee{\end{equation}}
\def\bea{\begin{eqnarray}}   \def\eea{\end{eqnarray}}
\def\bean{\begin{eqnarray*}} \def\eean{\end{eqnarray*}}
\def\bfr{\begin{flushright}} \def\efr{\end{flushright}}
\def\bc{\begin{center}}      \def\ec{\end{center}}
\def\bal{\begin{align}}      \def\eal{\end{align}}
\def\ba#1{\begin{array}{#1}} \def\ea{\end{array}}
\def\bd{\begin{description}} \def\ed{\end{description}}
\def\bv{\begin{verbatim}}    \def\ev{\end{verbatim}}

\def\lft{\left}                  \def\rgt{\right}
\def\la{{\langle}}               \def\ra{{\rangle}}

\def\Halmos{\hfill\vrule height10pt width4pt depth2pt \par\hbox to \hsize{}}
\def\pref#1{(\ref{#1})}
\def\lb#1{\label{#1}}

\def\qed{\raise1pt\hbox{\vrule height5pt width5pt depth0pt}}
\let\dpr=\partial

\let\==\equiv

\let\io=\infty
\let\0=\noindent

\def\*{{\hfill\break\null\hfill\break}}

\def\insertplot#1#2#3#4#5#6{%
\begin{figure}[ht]
\begin{center}
\includegraphics[bb=0 0 #1 #2, clip,  
width=#1pt, height=#2pt, 
clip=true, viewport=0 0 #1 #2]{#4.ps}
\caption{#5}
\end{center}
\end{figure}
}

\begin{document} 
\title{Interacting Fermions Picture for Dimer Models}
%\title{A new theoretical approach to Interacting Dimers}
%\title{Interacting Fermions Picture for Dimer Models}%

\author{P. Falco }
\affiliation{Department of Mathematics, 
California State University, Northridge,
CA 91330}

\begin{abstract}
Recent numerical results on classical dimers with weak aligning 
interactions have been  
theoretically justified via a Coulomb Gas
representation of  the 
height random variable. 
Here we propose a completely different representation, 
the Interacting Fermions Picture, which avoids some difficulties of the 
Coulomb Gas approach and provides a better
account of the numerical findings. 
Besides, we observe that   Peierls' argument explains  
the behavior of the system in the strong 
interaction case. 
\end{abstract}

\maketitle

\section{Introduction}
The lattice model of hard-core 
close-packed dimers 
is among the most fundamental %systems
in two-dimensional Statistical Mechanics. Not only 
it is exactly solvable in the narrow sense 
(the free energy and many correlations can be computed
\cite{Kas63, Lie67d, FiSte63, Ke97, Dub11a})   
but it also provides,  through equivalences, 
the exact solutions of several other models, including 
the nearest neighbor Ising model \cite{Kas63,Fis66},  and  some vertex models
at the so called {\it free fermion point} 
\citep{FW70,WuLi75}. 

Recently, especially in connection with a problem of Quantum Statistical Mechanics, 
\cite{RK88},  
several authors have been studying  the {\it classical} 
dimer model on a square lattice, modified by 
a weak aligning interaction. 
We will call it {\it interacting dimer model} (IDM). 
Since an exact solution of the 
IDM is not known, 
%(with an exception in \cite{HWKK96})  
our knowledge of the properties of this system %- quite a deep knowledge, in fact - 
rests entirely 
on the numerical analysis of  \cite{AJMPMT05,AIJMP06,TPAP07,Ot09};  
and on their theoretical interpretation 
via  the {\it Coulomb Gas approach}, (CGA). 

The CGA \cite{Kad78,dN83,Nie84} (see also \cite{Jac09})
applies to every  model for which one can define a  
{\it height random variable}; % as is the case of the IDM; 
and  it is based on the postulate that, in 
the scaling limit,  height correlations are equal to 
charge correlations of the free boson field.
This approach  has been very successful in explaining  
long range correlations in  
very many models of two-dimensional lattice 
 Statistical Mechanics. 
From a practical viewpoint, though, the conjectured 
scaling limit of the height correlation
has been difficult to substantiate (the best  
result in this direction is  \cite{FrSp81c}). This implies that, 
for example, in the case of the IDM,   
the  CGA has  not provided so far: a) an account of 
the staggered prefactors of dimer correlations; 
b) the dependence of the critical exponents in the 
coupling constant of the model.   

This Article proposes an alternative method for studying the IDM, 
that we call
{\it Interacting  Fermions Picture} (IFP). The basic idea is not new in Physics.  
It has been a standard tool in condensed matter theory for studying 
the 1+1-dimensional quantum models (see  \citep{So79,Gia04}). 
In classical two-dimensional Statistical Mechanics, 
it was first employed  in \cite{Spe00} to demonstrate the universality 
of the nearest neighbor Ising model under small, ``solvability  breaking'', 
perturbations;  subsequently
it was used to study the weak-universal properties 
of the Eight-Vertex and the Ashkin-Teller models \cite{GiMa04a,BFM10}. 
In relation to the IDM, 
fermion viewpoints have already been 
employed in \cite{FMS02, PLF07,YaKi12}; however, 
the IFP proposed here, as well as the results that we derive, 
appear to be completely new. 
The method 
is made of two steps: 
i) the IDM is re-casted into a lattice fermion field with a self-interaction; 
ii) the scaling limit of the lattice field 
is showed to be the Thirring model, which is interacting 
as well,  but which is also  exactly solvable.
The IFP solves the problems left open by  the 
CGA  because: a) it clarifies  the origin 
of  the staggering prefactors of the dimers correlations; 
b) it %the scaling limit argument is concrete, and 
provides the relationship between the 
correlations critical exponents and the coupling constant 
as a series of Feynman graphs.   
% \cite{BFM09a};
%
%Besides, %in the IFP,  
%via {\it bosonization} of the Thirring model, 
%one could always  recover   
%the free boson description that is typical of the CGA,
%without ever introducing the height variable of the lattice model.     
%For these reasons, i

Before  concluding,  we also provide (by a different argument) 
a theoretical explanations of the
numerical results in the strong interaction case. 
 
\section{Definitions and Results }
Consider a finite box $\L$ of the infinite square lattice.
A {\it dimer configuration}, $\o$, is a collection of 
dimers covering the edges of $\L$ 
with the constraint that every vertex of $\L$ is covered by one, 
and only one, dimer.
%If we assign to the edge $d$ of $\L$ the activity $z_d$, 
The partition 
function of the interacting dimers model (IDM) is 
\be\lb{pf}
Z_\l(\L)=\sum_{\o}\exp\Big\{\l\sum_{d, d'\in \o} v(d,d')\Big\}%\prod_{d\in \o} z_d
\ee
where: $\l$ is the dimers coupling constant;  
$v(d,d')$ is a  two body dimer interaction; the first sum is over all the 
dimer configurations; the second sum is over any pair of
dimers in the  configuration $\o$. %of the configuration $\o$: 
In \cite{AIJMP06} $\l v(d,d')$ is 
a special, nearest neighbor, aligning interaction. In this work 
we only  assume that: $v(d,d')$  is  zero unless $d$ and $d'$ are both 
horizontal or both vertical; that it is 
invariant under $\p/2$-rotations and  under lattice translations;  
and that $|v(d,d')|$ has exponential decay in the distance 
between $d$ and $d'$.
The ``non-interacting'', exactly solvable,  dimer model is the case $\l=0$
\cite{Kas63, Lie67d, FiSte63, Ke97, Dub11a}.

Our main result  
is the evaluation of  correlation critical exponents
of local bulk observables, for small  $|\l|$. 
A natural observable to consider is the {\it dimer occupancy} $\n_d(\o)$, 
which is equal to 1 if the dimer $d$ is present in $\o$, and 
zero otherwise. 
Consider the horizontal dimers $d=\{0,\bfe_0\}$ and $d'=\{\xx, \xx+\bfe_0\}$, 
for $\bfe_0=(1,0)$ and $\xx=(x_0, x_1)$. The IFP provides the 
large-$|\xx|$ formula  
\begin{align}\lb{ddc}
\la \n_d\n_{d'}\ra -\la \n_d\ra \la \n_{d'}\ra
&\sim
 (-1)^{x_0+x_1} c 
\frac{x_0^2-x_1^2}{(x_0^2+x_1^2)^2}
\notag\\
&+(-1)^{x_0} c_- 
\frac{1}{(x_0^2+x_1^2)^{\kappa_-}}\;,
\end{align}
for a $\l,  v$-dependent {\it critical exponent} $\kappa_-=1+O(\l)$
and {\it staggering prefactors} $(-1)^{x_0+x_1} $ and $(-1)^{x_0}$. 
 % All the other 
%sub-leading terms are neglected.  This formula  
For $\l=0$ this result coincides with the exact solution: see (7.12) and (7.20) of \cite{FiSte63};  and it 
agrees with 
the numerical simulations  for small positive $\l$: 
see (51) and (52) of \cite{AIJMP06}
 
The critical exponent $\kappa_-$ is non-universal, because 
it does depend on $\l$ and $v(d,d')$. 
What is expected to be 
universal, instead,  %(and so is a useful prediction for experimental physics) 
is the relationship among  critical exponents of different 
observables.    
It is instructive  to study a second observable, then. 
The authors of \cite{AIJMP06} considered the 
monomer correlation%. 
%We could describe this correlation in the IFP as well
; however,
being it  equivalent to a {\it non-local }
fermion correlation, 
the derivation of the  scaling limit in the IFP is, 
at the present time,  not more transparent 
than in the  CGA.
We consider instead a different observable, the {\it diagonal dimer}. It 
consists in a  pair of monomers in positions $\{\xx, \xx+\bfe\}$, where 
$\bfe=(1,1)$; see Fig.\ref{f1}. 
Since such a dimer is not allowed in the hard-core, close-packed 
 configurations $\o$, 
we define its ``correlation''  
as it is done for the monomer observable, 
i.e. in terms of lattice defects: 
$$
\la \n_d\ra =\lim_{\L\to \io} \frac{Z_\l (\L-d)}{Z_\l (\L)}
\quad
\la \n_d \n_{d'}\ra =\lim_{\L\to \io} \frac{Z_\l (\L-(d\cup d'))}{Z_\l (\L)}\;.
$$
For $d=\{0,\bfe\}$ and $d'=\{\xx, \xx+\bfe\}$, the IFP
gives the large-$|\xx|$ formula
\begin{align}\lb{ddc2}
\la \n_d\n_{d'}\ra -\la \n_d\ra \la \n_{d'}\ra
&\sim
c_+
\frac{(-1)^{x_0+x_1}-1}{(x_0^2+x_1^2)^{\kappa_+}}\;,
\end{align}
for a new  $\l, v$-dependent {\it critical exponent} $\kappa_+=1+O(\l)$
and for a {\it staggering prefactor} $(-1)^{x_0+x_1}-1$. 
The universal formula that relates $\kappa_+$ to $\kappa_-$ is 
peculiar  of the models with central charge $c=1$ and was originally 
discovered (in a different model) by  
 Kadanoff \cite{Kad77}:
%for any (short range) choice of $v$ and any 
%(small enough) $|\l|$ 
\be\lb{kf}
\kappa_+\cdot \kappa_-=1\;.
\ee
In the next sections we  will derive our main results:
\pref{ddc}, \pref{ddc2} and \pref{kf}.  
%We remark 
As by-product, we will obtain   
a Feynman graphs representation of the expansion of $\kappa_\pm$ 
in powers of $\l$. For example, at first order
(for $\hat v$ the Fourier transform of the interaction $v$ 
--in \cite{AIJMP06} $\hat v(\kk)=\cos k_0+ \cos k_1$)
\be\lb{cecp}
\kappa_-=1+\frac{8}{\p}\lft[\hat v(\p,\p)-\hat v(0,\p)\rgt]\l +O(\l^2)\;;
\ee
 hence, according to 
the sign of $\lft[\hat v(\p,\p)-\hat v(0,\p)\rgt]\l$ %(in \cite{AIJMP06} $\hat v(\p, \p)=-2$)
either the former or the latter term 
in \pref{ddc} is dominant at large distances.
Note that the CGA was successfully  used to justify the appearance  
of critical exponents $\kappa_\pm$ which satisfy \pref{kf}
(see, for example, pt.1 and pt.5 on page 7 of \cite{PLF07}).
However,  the CGA does not provide   
the staggering prefactors of \pref{ddc} and \pref{ddc2}; nor 
does it provide  
any  relationship between $\kappa_\pm$ and $\l$ such as \pref{cecp}.
%Instead,
%On the contrary
\insertplot{140}{120}
{}{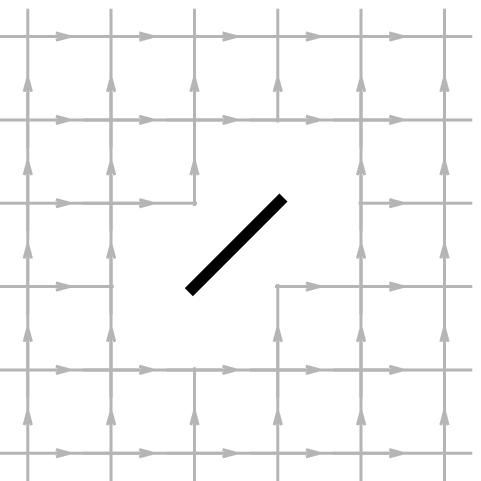}{Example of a diagonal dimer. Its presence in the graph generates 
a new face of clockwise oddness $-1$: thus, as opposed to the case of the 
monomer defect \cite{FiSte63}, the fermion representation of 
the diagonal dimer $\{\xx,\xx+\bfe\}$ remains local and is $(-1)^{x_1+1}i \ps_{\xx} \ps_{\xx+\bfe}$.  
\lb{f1} }{0} 
\section{Interacting Fermions Picture}
When $\l=0$ the dimer model is equivalent to a lattice fermion 
field without interaction. 
Namely
\be\lb{pf0}
Z_{0}(\L)=\int\! D\ps\; \exp\Big\{-\frac12\sum_{\xx, \yy} K_{\xx, \yy} \ps_\xx \ps_\yy\Big\}
%=\Pfa_{\xx, \yy} K_{\xx, \yy}
\ee
where: $\{\ps_\xx:\xx\in \L\}$ are Grassmann variables and  $D\ps$ indicates the 
integration w.r.t. all of them;  $K_{\xx, \yy}$ is the {\it  Kasteleyn matrix} 
that  can be chosen to be such that 
$$
\sum_\yy K_{\xx, \yy} \ps_\yy
=\sum_{\s=\pm1} \s(\ps_{\xx+\s\bfe_0}
+
i \ps_{\xx+\s\bfe_1}) 
$$
with $\bfe_0=(1,0)$ and $\bfe_1=(0,1)$. 
%Notice that 
%$K$ is a discrete version of $2(\dpr_0+i \dpr_1)$. 
\pref{pf0} is  the partition function of
a free {\it Majorana fermion field}, i.e. a Grassmann-valued 
Gaussian field  with moment 
generator 
$$
\la e^{\sum_\xx\ps_\xx\h_\xx}\ra_{0} = e^{-\frac12\sum_{\xx,\yy}S(\xx-\yy)\h_\xx\h_\yy}
$$  
where: the $\h_\xx$'s are other Grassmann variables; 
$S$, the covariance, is the inverse Kasteleyn 
matrix  
$$
S(\xx)=\la \ps_\xx \ps_0\ra_0=\frac12\int_{-\p}^\p\!\frac{d p_0}{2\p}
\int_{-\p}^\p\!\frac{d p_1}{2\p}\; \frac{e^{i p_0 x_0+i p_1 x_1}}
{i\sin p_0-\sin p_1}\;.
$$
The Fourier transform 
of $S$ is singular at four {\it Fermi momenta}: 
$\pp_{+,0}=(0,0)$, $\pp_{+,1}=(\p,\p)$, $\pp_{-,0}=(0,\p)$ and
$\pp_{-,1}=(\p,0)$.
Therefore, in view of the scaling limit, it is convenient  to decompose 
$$
\ps_\xx=\sum_{\o=\pm\atop s=0,1} i^s e^{i \pp_{\o,s}\xx}\ps_{\xx,\o,s}
$$
where $\ps_{\xx,\o,s}$ are four {\it independent} 
Majorana fields with large-$|\xx|$ covariances
\begin{align}
\la\ps_{\xx,\o,s} \ps_{0,\o,s}\ra_0 \;
&\sim\;
\frac12\int_{-\io}^\io\!\frac{d p_0}{2\p}\int_{-\io}^\io\!
\frac{d p_1}{2\p}\; \frac{e^{i p_0 x_0+i p_1 x_1}}
{ip_0-\o p_1}
\notag\\
&=\frac{1}{4\p}\frac1 {x_0+i\o x_1}\;.
\end{align}
This decomposition already appeared  in \cite{FMS02}
for studying  
the free case, which is exactly solvable. Instead
here we are preparing for the application to the interacting 
case and thus we also introduce  Dirac spinors
$\ps^+_\xx=(\ps^+_{\xx,+},\ps^+_{\xx,-} )$ and 
$\ps_\xx=(\ps_{\xx,+},\ps_{\xx,-} )^T$ for
\be\lb{drf}
\ps^+_{\xx,\o}=\frac{\ps_{\xx,\o,0}+i\ps_{\xx,\o,1}}{\sqrt2}
\quad
\ps_{\xx,\o}=\frac{\ps_{\xx,\o,0}-i\ps_{\xx,\o,1}}{\sqrt2}\;,
\ee
with translational invariant covariances
\begin{align}
\la \ps_{\xx,\o}&\ps_{0,\o'}\ra_0=0,\qquad
\la \ps^+_{\xx,\o}\ps^+_{0,\o'}\ra_0=0\;,
\notag\\
&\la \ps^+_{\xx,\o}\ps_{0,\o'}\ra_0\;\sim\;
\frac{\d_{\o,\o'}}{4\p}\frac1 {x_0+i\o x_1}\;.
\end{align}
%These asymptotic formulas coincide with the corresponding ones in the
%Quantum Field Theory of a free Dirac field 
%in dimension two (for a suitable choice of the gamma matrices). 
If we now let $\l\neq 0$, by power series expansion in $\l$
one can verify that \pref{pf} becomes
\be\lb{pf1}
Z_{\l}(\L)=\int\! D\ps\; 
\exp\Big\{-\frac12\sum_{\xx, \yy} K_{\xx, \yy} \ps_\xx \ps_\yy+V(\l v, \ps)\Big\}
\;,
\ee 
where $V(\l v, \ps)$ is a sum of even monomials in the $\ps$'s of 
order bigger than two. %,
%%
%$$
%V(\l v, \ps)=\l \sum_{d, d'}\F_d \F_{d'} v(d,d')+ O(\l^2, \F^3)  
%$$
%%
%with $\F_d=K_{\xx,\xx'}\ps_{\xx}\ps_{\xx'}$, 
%if $d$ is the lattice bond  $\{\xx, \xx'\}$. 
%Of the exact form 
%of $V$, what is important is a set of symmetries given below.
It is not difficult to see that 
the dimer correlation in the l.h.s. of \pref{ddc} becomes,
in terms of  Dirac fermions \pref{drf} 
and up to terms with  faster decays
\begin{align}\lb{ddc3}
&\la \ps_0 \ps_{\bfe_0} \ps_{\xx} \ps_{\xx+\bfe_0}\ra - 
\la \ps_0 \ps_{\bfe_0}\ra \la \ps_{\xx} \ps_{\xx+\bfe_0}\ra
\notag\\
&\sim
4(-1)^{x_0+x_1}
\sum_\o\la \ps^+_{0,\o} \ps_{0,\o}; \ps^+_{\xx,\o}\ps_{\xx, \o} \ra^T
\notag\\
&+
4(-1)^{x_0}\sum_\o 
\la\ps_{0,\o}^+\ps_{0,-\o};\ps_{\xx,-\o}^+\ps_{\xx,\o}\ra^T\;,
\end{align}
where the label $T$ indicates a truncated correlation. 
 In the same way, 
the diagonal dimer correlation in the l.h.s. of \pref{ddc2} becomes 
\begin{align}
&-(-1)^{x_1}\lft[\la \ps_0 \ps_{\bfe} \ps_{\xx} \ps_{\xx+\bfe}\ra - 
\la \ps_0 \ps_{\bfe}\ra \la \ps_{\xx} \ps_{\xx+\bfe}\ra\rgt]
\notag\\
&\sim- 8\lft[(-1)^{x_0+x_1}-1\rgt]
\la \ps^+_{0,+} \ps^+_{0,-}; \ps_{\xx, -}\ps_{\xx,+} \ra^T\;.
\end{align} 
In the next section, 
by a  Renormalization Group argument, we will 
explain  why, in the evaluation of the large distance decay of the 
correlations, 
it is correct to replace 
the interacting  fermion field \pref{pf1} with the massless Thirring model. 
Assuming for the moment this crucial fact, we only need to borrow the  
exact solutions for the Thirring model correlations 
\cite{Jo61,Kla64,Ha67} (see also \cite{Fa12b}):
%The latter operation cost other error terms, with faster large distance decay. 
%
\begin{align}
\la \ps^\dagger_{\o}(0) \ps_{\o}(0); \ps^\dagger_{\o}(\xx)\ps_{\o}(\xx) \ra^T
&=c
\frac{x_0^2-x_1^2}{(x_0^2+x_1^2)^2}
\notag\\
\la \ps^\dagger_{+}(0) \ps^\dagger_{-}(0); \ps_{-}(\xx)\ps_{+}(\xx) \ra^T
&=
\frac{c_+}{(x_0^2+x_1^2)^{\kappa_+}}
\notag\\
\la\ps_{\o}^\dagger(0)\ps_{-\o}(0);\ps_{-\o}^\dagger(\xx)\ps_{\o}(\xx)\ra^T
&=
\frac{c_-}{(x_0^2+x_1^2)^{\kappa_-}}
\end{align}
where the critical exponents $\kappa_\pm$ are
$$
\kappa_+=\frac{1+\frac{\l_T}{4\p}}{1-\frac{\l_T}{4\p}}
\qquad
\kappa_-=\frac{1-\frac{\l_T}{4\p}}{1+\frac{\l_T}{4\p}}
$$
and $\l_T$ is a parameter of the Thirring model: at first order 
$\l_T=-16\lft[\hat v(\p,\p)-\hat v(0,\p)\rgt]\l+O(\l^2)$ (see next section).  
The derivation of  \pref{ddc}, \pref{ddc2}, \pref{kf} is complete. 
%In fact, the RG argument of the next section gives a precise
%series representation of $\l_T$ as powers of $\l$.
%Notice that 
%$c_+=\frac1{2\p^2}+O(\l)$, 
%$c(\l)=\frac1{2\p^2}+O(\l)$ and $c_0=\frac1{2\p^2}+O(\l)$.  
%
\section{RG Analysis}
We follow  Wilson's RG scheme in the version due to Gallavotti 
\cite{Ga85}. Integrating out the large momentum scales, one obtains 
an effective interaction 
\begin{align}
&\sum_n\sum_{\o_1,\ldots, \o_n\atop s_1,\ldots, s_{2n}}\frac{i^{s_1+\cdots s_{2n}}}{(2\p)^{4n-1}} 
\int d\kk_1\cdots d\kk_{2n}\; 
\hp_{\kk_1, \o_1, s_1}\cdots \hp_{\kk_{2n}, \o_{2n}, s_{2n}} 
\notag\\
&\cdot\d\Big(
\sum_{j=1}^{2n}\kk_j+\sum_{j=1}^{2n}\pp_{\o_j,s_j}\Big)
\hw_{2n}(\kk_2+\pp_{\o_2,s_2},\ldots,\kk_{2n}+\pp_{\o_{2n},s_{2n}})\;,
\notag
\end{align}
where $\hw_{2n}$'s are series of Feynman graphs. 
Some symmetries are of crucial importance. 
For $R(k_0,k_1)=(k_1,-k_0)$,
$\th (k_0,k_1)=(k_1,k_0)$ and $\t (k_0,k_1)=(k_0,k_1+\p)$, 
we find
\begin{align}\lb{sym}
&\hw_{2m}(\t\kk_2, \ldots, \t\kk_{2m})=(-i)^m\hw_{2m}(\th\kk_2, \ldots, \th\kk_{2m})
\notag\\
&\hw_{2m}(R\kk_2, \ldots, R\kk_{2m})=(-i)^m\hw_{2m}(\kk_2, \ldots, \kk_{2m})
\notag\\
&\hw_{2m}(\kk_2, \ldots, \kk_{2m})^*=i^m\hw_{2m}(\th\kk_2, \ldots, \th\kk_{2m})
\;.
\end{align}
From power counting, 
there are two possible local, marginal terms: 
a quartic term,  that requires the renormalization 
of the coupling constant $\l$; a quadratic term, responsible for a
field renormalization. By \pref{sym} they are: 
\begin{align}\lb{4}
24\hw_4(\pp_{+,1},\pp_{-,0},\pp_{-,1})
\sum_{\xx}\ps_{\xx,+}^+\ps_{\xx,+}^-\ps_{\xx,  -}^+\ps_{\xx, -}^- 
\end{align}
and 
\begin{align}\lb{2}
2\lft[-i\frac{\dpr \hw_2}{\dpr k_0}(\pp_{+,0})
-\frac{\dpr \hw_2}{\dpr k_1}(\pp_{+,0})\rgt]
\sum_{\xx,\o} 
\ps_{\xx, \o}^+ \dpr_{\o}\ps_{\xx, \o}^- \;,
\end{align}
where $\dpr_\o$ is the  Fourier transform of $ik_0-\o k_1$. Again by \pref{sym},
the prefactors in \pref{4} and \pref{2} are real. 
Instead, there are no local, relevant terms: 
the only possible one, a quadratic term without derivatives, 
i.e. a {\it mass term}, cannot be generated by  conservation of
 total momentum. 
These facts imply  
that \pref{pf1}, with parameter $\l$, 
equals the massless 
Thirring model, with parameter 
$\l_T$, up to  terms which are
irrelevant and thus  cannot modify \pref{kf}
(although they do contribute to  the relationship between $\l$ and $\l_T$
as series of Feynman graphs).

\section{Strong Interaction}
In the opposite case of {\it strong} dimer interaction, i.e. $|\l|\gg1$, 
the numerical findings of \cite{AIJMP06} indicate 
the existence of four different ``columnar'' phases.
Also this outcome can be theoretically explained% by a method that is more robust than 
%the CGA
: not by the IFP this time, but rather by the classical  Peierls' argument.  However, as opposed 
to the weak interaction case, the outcome does  depend on the choice of %the interaction 
$v(d,d')$: for definiteness, we discuss here the choice
in \cite{AIJMP06}, which assigns an energy  $-\l<0$
per each plaquette displaying one of the two dimer arrangements 
in Fig.\ref{f2}.
Decorate a dimer 
configuration with wiggly lines as indicated in Fig.\ref{f2}:
a plaquette is  ``good''  if it contains 2 wiggly 
lines; otherwise, 
 the plaquette is  ``bad''. Note that:  a) dimer configurations on 
nearest neighbor good plaquettes must correspond to the same columnar ground state;  
b) the probability of a bad plaquettes occurrence is  
dumped, at least, by a factor $e^{-\frac\l4}$ per plaquette. 
Therefore
one can apply the  
Peierls' argument to the ``contours'' of bad plaquettes
(see \cite{HeiLi79,FILS80} or the review \cite{Bi09})  
and show that, for $\l$ positive and large,  the number of different phases 
coincides with  the number of ground states.
\insertplot{140}{50}
{}{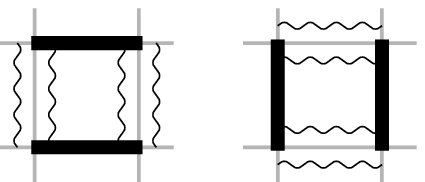}{ 
Draw four wiggly lines for each plaquette containing two facing dimers.  
An equivalent, but completely local way of evaluating  the total energy of a dimer 
configuration is then by assigning an energy of $-\frac\l4$ per wiggly line. \lb{f2} }{0} 
\section{Conclusion}
In the case of weak short ranged interactions, 
we have showed that the IFP %Interacting Fermion Picture 
provides a detailed account of the 
numerical findings of \cite{AIJMP06} --plus  some new predictions. 
This method should also work, 
with possibly different outcomes, for  triangular and hexagonal lattices; and
for two or more interacting copies of dimer models.
Besides, the IFP should be applicable to the Six-Vertex model, which
is equivalent to dimers on a square lattice with a staggered interaction.
Including the results on Ashkin-Teller,  Eight-Vertex and XYZ quantum chain \cite{BFM10}
the IFP seems quite an effective way  for dealing with two dimensional lattice 
critical models with central charge $c=1$.    
It might be possible that the IFP be applicable to $c<1$ models 
(for example via  ideas in  \cite{dN83}).
%The answer is apparently positive, because of the triple equivalence (see Sec. 2.1.1)
%$q-$states Potts models /  close-packed  $O(n)$ loop 
%models /  6V model. Anyways the difficulty here is in the bc, that
%are so relevant  to lower
%the central charge value from the $c=1$ of 6V to $c<1$. This problem is 
%clearly very difficult, 
%but also very important: any rigorous result about correlations in such integrable models  
%would be new and long awaited.  

In the  case of strong aligning interactions, we have explained  the 
numerical findings of \cite{AIJMP06} by  Peierls' argument.  

\section{Acknowledgments}
During the  elaboration of the ideas and  the 
the preparation of the work 
I benefited of several discussions with 
T. Spencer,  M. Biskup and F. Bonetto.

\bibliographystyle{apsrev4-1}
\bibliography{widl.bib}

\end{document}